\documentclass[12pt]{article}
\usepackage[numbers,sort&compress]{natbib}
\usepackage[utf8]{inputenc}
\usepackage[english]{babel}
\usepackage{fullpage}
\usepackage{amsmath}
\usepackage{amssymb}
\usepackage{MnSymbol}
\usepackage{graphicx}
\usepackage{wrapfig}
\usepackage{setspace}
\usepackage{xr-hyper}
\usepackage[pdftex,bookmarks=true]{hyperref}
\hypersetup{
   colorlinks,
   citecolor=blue,
   filecolor=black,
   linkcolor=blue,
   urlcolor=blue
}
\usepackage{cleveref}
\allowdisplaybreaks

\def\pdotq{p \cdot q}

\newcommand{\bv}[1]{{\mathbf{#1}}}

\def\llangle{\left\langle}
\def\rrangle{\right\rangle}
\newcommand{\ea}[1]{{\llangle #1 \rrangle}}

\newcommand\pubdate{\today}

\def\Title#1{\begin{center} {\Large #1 } \end{center}}
\def\Author#1{{ \sc #1}}
\def\Address#1{{ \it #1}}

\newcommand\pubblock{\rightline{\begin{tabular}{l}  \\ % Author's note number [if you need to add one] goes here
         \pubdate  \end{tabular}}}
\newenvironment{Abstract}{\begin{quotation}  }{\end{quotation}}
\newenvironment{Presented}{\begin{quotation} \begin{center} 
             PRESENTED AT\end{center}
      \begin{center}\begin{large}}{\end{large}\end{center} \end{quotation}}
\begin{document}
\begin{titlepage}
 \pubblock
\vfill
\Title{The Compton amplitude and nucleon structure functions in lattice QCD}
\vfill
\Author{K.~U.~Can${}^{a,\ast}$, M.~Batelaan${}^a$, A.~Hannaford-Gunn${}^a$, R.~Horsley${}^b$, Y.~Nakamura${}^c$, H.~Perlt${}^d$, P.~E.~L.~Rakow${}^e$, G.~Schierholz${}^f$, H.~St\"{u}ben${}^g$, R.~D.~Young${}^a$, and J.~M.~Zanotti${}^a$} \\

\noindent \hspace{-3mm} 
\Address{${}^a$ CSSM, Department of Physics, The University of Adelaide, Adelaide SA 5005, Australia.} \\
\Address{${}^b$ School of Physics and Astronomy, University of Edinburgh, Edinburgh EH9 3JZ, UK.} \\
\Address{${}^c$ RIKEN Center for Computational Science, Kobe, Hyogo 650-0047, Japan.} \\
\Address{${}^d$ Institut f\"{u}r Theoretische Physik, Universit\"{a}t Leipzig, 04103 Leipzig, Germany.} \\
\Address{${}^e$ Theoretical Physics Division, Department of Mathematical Sciences, University of Liverpool, Liverpool L69 3BX, UK.} \\
\Address{${}^f$ Deutsches Elektronen-Synchrotron DESY, Notkestr. 85, 22607 Hamburg, Germany.} \\
\Address{${}^g$ Regionales Rechenzentrum, Universit\"{a}t Hamburg, 20146 Hamburg, Germany.} \\
\Address{${}^\ast$ Speaker}

\vfill
\begin{Abstract}
\noindent
The structure of hadrons relevant for deep-inelastic scattering are completely characterised by the Compton amplitude. A direct calculation of the Compton amplitude in a lattice QCD setup provides a way to accessing the structure functions, circumventing the operator mixing and renormalisation issues of the standard operator product expansion approach. \\

\vspace{-3mm} \noindent
In this contribution, we focus on the QCDSF/UKQCD Collaboration's advances in calculating the forward Compton amplitude via an implementation of the second-order Feynman-Hellmann theorem. We highlight our progress in investigating the moments of nucleon structure functions. 
\end{Abstract}
\vfill
\begin{Presented}
DIS2023: XXX International Workshop on Deep-Inelastic Scattering and
Related Subjects, \\
Michigan State University, USA, 27-31 March 2023 \\
     \includegraphics[width=9cm]{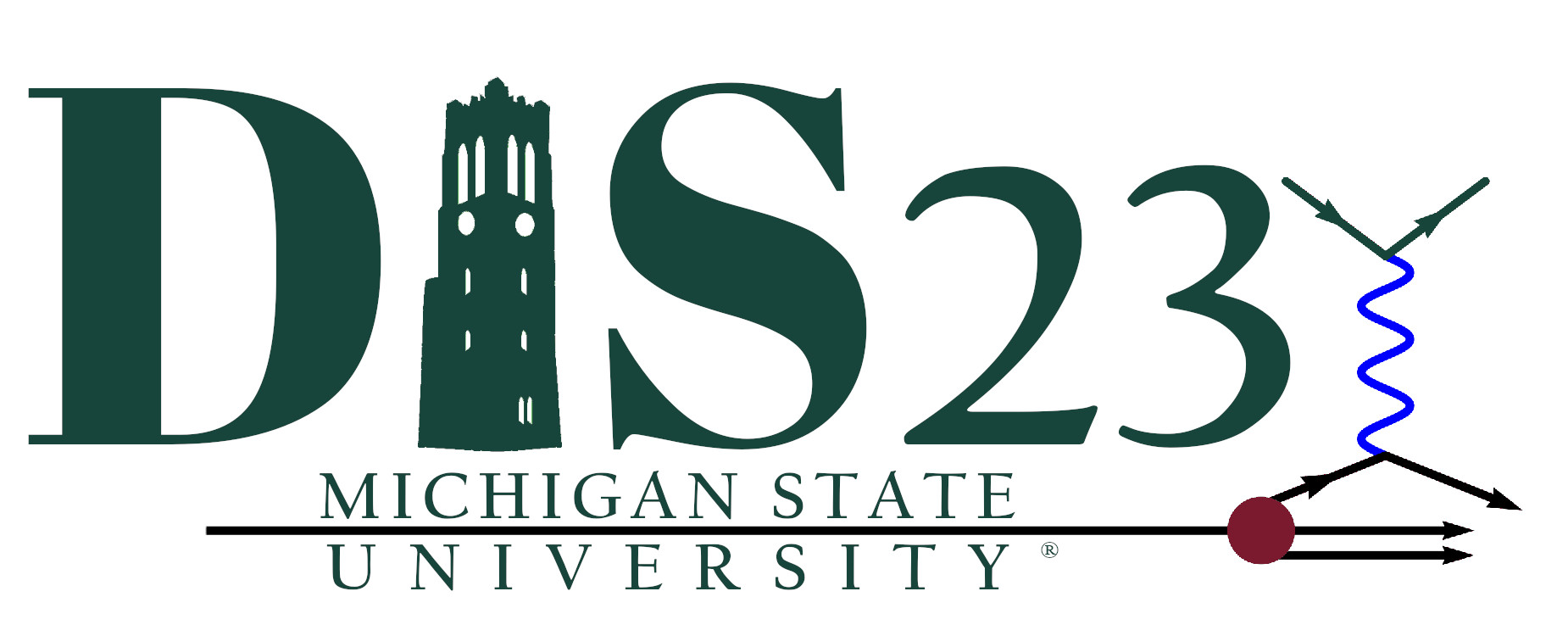}
\end{Presented}
\vfill
\end{titlepage}

\section{Introduction}
Understanding the internal structure of the nucleon is a long standing and intriguing subject in the field of hadron physics. Structure functions are the essential components that encode the dynamics of the quark and gluon degrees of freedom at short distances. 
Calculating the structure functions from first principles poses several challenges for lattice QCD practitioners. Traditionally, lattice calculations make use of the operator product (OPE) expansion. However, it is known that in the OPE approach, contributions of leading-twist operators are inseparably connected with the contributions from operators of higher twist, due to operator mixing and renormalisation~\cite{Martinelli:1996pk}. \\

\vspace{-3mm} \noindent
Here we describe an approach that is being pursued by the QCDSF/UKQCD Collaboration, which is to directly calculate the forward Compton amplitude on the lattice in the space-like region~\cite{PhysRevLett.118.242001}. By working with the physical amplitude, the operator mixing and renormalization issues, and the restriction to light-cone operators are circumvented. Once we are able to determine the Compton amplitude sufficiently accurately, we can expect to estimate the power corrections in structure functions, i.e. quantify the target mass corrections and estimate the contributions from higher-twist operators, which could be useful for global PDF analyses. This contribution is partially based on Refs.~\cite{PhysRevD.102.114505,QCDSF:2022ncb}. Although our focus is the Compton amplitude in forward kinematics, this approach is applicable to off-forward kinematics enabling an investigation of the generalised parton distributions~\cite{Alec:2021lkf}.

\section{The Compton tensor and the moments of structure functions}
We start from the forward Compton amplitude described by the time ordered product of electromagnetic currents sandwiched between nucleon states, 
\begin{align} 
    \label{eq:compamp}
    T_{\mu\nu}(p,q) =& \int d^4z\, e^{i q \cdot z} \rho_{s s^\prime} \ea{p,s^\prime \left| 
    \mathcal{T}\left\{ \mathcal{J}_\mu(z) \mathcal{J}_\nu(0) \right\} \right|p,s},
\end{align}
where $p$ is the momentum and $s$ is the spin of the nucleon, $q$ is the momentum of the virtual photon, and $\rho$ is the polarisation density matrix. For parity-conserving processes that involve conserved currents the spin-averaged Compton tensor is parametrised in terms of two Lorentz-invariant scalar functions, $\mathcal{F}_1$, and $\mathcal{F}_2$ as follows
\begin{equation} \label{eq:compamp_tensor}
    T_{\mu\nu}(p,q) = \left( -g_{\mu\nu} + \frac{q_\mu q_\nu}{q^2} \right) \mathcal{F}_1(\omega,Q^2) + \hat{P}_\mu \hat{P}_\nu \frac{\mathcal{F}_2(\omega,Q^2)}{\pdotq}
\end{equation}
where $\hat{P}_\mu \equiv p_\mu - q_\mu (\pdotq/q^2)$, with $Q^2 = -q^2$ and $\omega = 2 (\pdotq)/Q^2$. \\

\vspace{-3mm} \noindent 
The Compton structure functions are related to the corresponding ordinary structure functions via the optical theorem, $\operatorname{Im}\mathcal{F}_{1,2}(\omega,Q^2) = 2\pi F_{1,2}(x,Q^2)$. Making use of analyticity, crossing symmetry and the optical theorem, we can write dispersion relations for $\mathcal{F}$,
\begin{align}
    \label{eq:compomega12} 
    \overline{\mathcal{F}}_1(\omega,Q^2)= 2\omega^2 \int_0^1 dx \frac{2x \, F_1(x,Q^2)}{1-x^2\omega^2-i\epsilon}, \quad
    \mathcal{F}_2(\omega,Q^2)= 4\omega \int_{0}^1 dx\, \frac{F_2(x,Q^2)}{1-x^2\omega^2-i\epsilon},
\end{align}
where $\overline{\mathcal{F}}_1(\omega,Q^2) = \mathcal{F}_1(\omega,Q^2)-\mathcal{F}_1(0,Q^2)$ is the once subtracted structure function. Although we are only concerned with subtracting it away, understanding the subtraction function is an interesting subject in itself. Related discussions on the subtraction function can be found in~\cite{Walker-Loud:2012ift,Hagelstein:2020awq,Lozano:2020qcg,HannafordSankey:2021lat}. \\

\vspace{-3mm} \noindent
Expanding the integrands in \Cref{eq:compomega12} at fixed $Q^2$ as a geometric series, the Compton structure functions can be expressed as infinite sums over the Mellin moments of the inelastic structure functions,
\begin{align} 
    \label{eq:ope_moments1}
    \overline{\mathcal{F}}_1(\omega,Q^2)&=\sum_{n=1}^\infty 2\omega^{2n} M^{(1)}_{2n}(Q^2), &&\text{with} \; M^{(1)}_{2n}(Q^2)= 2\int_0^1 dx\, x^{2n-1} F_1(x,Q^2), \\
    \label{eq:ope_moments2}
    \mathcal{F}_2(\omega,Q^2)&= \sum_{n=1}^\infty 4\omega^{2n-1} M^{(2)}_{2n}(Q^2), &&\text{with} \; M^{(2)}_{2n}(Q^2)= \int_{0}^1 dx\,x^{2n-2} F_2(x,Q^2).
\end{align}
We note that the physical moments $M_{2n}$ that appear in \Cref{eq:ope_moments1,eq:ope_moments2} are dominated by their leading-twist contributions, i.e. the moments of PDFs, at asymptotically large $Q^2$. \\ 

\vspace{-3mm} \noindent
Finally, the unpolarised Compton structure functions $\mathcal{F}_1$ and $\mathcal{F}_2$ are accessed from the Compton tensor via,
\begin{align}
    \label{eq:compF1}
    \mathcal{F}_1(\omega, Q^2) &= T_{33}(p,q), &&\text{for} \, \mu=\nu=3 \, \text{and} \, p_3=q_3=0, \\
    \label{eq:compF2}
    \frac{\mathcal{F}_2(\omega,Q^2)}{\omega} &= \frac{Q^2}{2 E_N^2} \left[ T_{00}(p,q) + T_{33}(p,q) \right],
    && \text{for} \, \mu=\nu=0 \, \text{and} \, p_3=q_3=q_0=0.
\end{align}

\section{Feynman-Hellmann technique}
Our implementation of the second order Feynman-Hellmann method is presented in detail in~\cite{PhysRevD.102.114505}. Here, we briefly summarise its main aspects. To start, we modify the fermion action with the following perturbing term,
\begin{equation}\label{eq:fh_perturb}
    S(\lambda) = S + \lambda \int d^3z (e^{i \bv{q} \cdot \bv{z}} + e^{-i \bv{q} \cdot \bv{z}}) \mathcal{J}_{\mu}(z) ,
\end{equation}
where $\lambda$ is the strength of the coupling between the quarks and the external field, $\mathcal{J}_{\mu}(z) = Z_V \bar{q}(z) \gamma_\mu q(z)$ is the electromagnetic current coupling to the quarks, $\bv{q}$ is the external momentum inserted by the current and $Z_V$ is the renormalization constant for the local electromagnetic current, which has been determined in Ref~\cite{Constantinou:2014fka}. The perturbation is introduced on the valence quarks only, hence only quark-line connected contributions are taken into account in this work. For the perturbation of valence and sea quarks see~\cite{Chambers2015}. \\

\vspace{-3mm} \noindent
The main strategy to derive the relation between the energy shift and the matrix element is to work out the second-order derivatives of the two-point correlation function with respect to the external field from two complementary perspectives. Differentiating the energy of the perturbed nucleon correlator, $G^{(2)}_\lambda(\bv{p};t) \simeq A_\lambda(\bv{p}) e^{-E_{N_\lambda}(\bv{p}) t}$,  one finds a distinct temporal signature for the second-order energy shift, and by matching it to the expression, coming from a direct evaluation of the correlator, one arrives at the desired relation between the energy shift and the matrix element describing the Compton amplitude,
\begin{equation} \label{eq:secondorder_fh}
    \left. \frac{\partial^2 E_{N_\lambda}(\bv{p})}{\partial \lambda^2} \right|_{\lambda=0} = - \frac{T_{\mu\mu}(p,q) + T_{\mu\mu}(p,-q)}{2 E_{N}(\bv{p})},
\end{equation}
where $T$ is the Compton amplitude defined in \Cref{eq:compamp}, $q=(0,\bv{q})$ is the external momentum encoded by \Cref{eq:fh_perturb}, and $E_{N_\lambda}(\bv{p})$ is the nucleon energy at momentum $\bv{p}$ in the presence of a background field of strength $\lambda$. This expression is the principal relation that we use to access the Compton amplitude and hence the Compton structure functions as in \Cref{eq:compF1,eq:compF2}. For a more detailed derivation, see~\cite{PhysRevD.102.114505,QCDSF:2022ncb}. \\

\vspace{-3mm} \noindent
The implementation of the Feynman-Hellmann theorem described above effectively inserts an external current on to a quark line by computing its propagator with the perturbed quark action \Cref{eq:fh_perturb}. When both currents are inserted onto the $u$-quarks or the $d$-quark, we evaluate the ``$uu$'' or ``$dd$'' contributions to the Compton structure functions, respectively.

\section{Selected results and discussion} \label{sec:res}
Our simulations are carried out on QCDSF/UKQCD-generated $2+1$-flavour gauge configurations. Two ensembles are used with volumes $V=[32^3 \times 64, 48^3 \times 96]$, and couplings $\beta=[5.50, 5.65]$ corresponding to lattice spacings $a=[0.074, 0.068] \, {\rm fm}$, respectively. Quark masses are tuned to the $SU(3)$ symmetric point where the masses of all three quark flavours are set to approximately the physical flavour-singlet mass, $\overline{m} = (2 m_s + m_l)/3$~\cite{Bietenholz:2010jr,Bietenholz:2011qq}, yielding $m_\pi \approx [470, 420] \, {\rm MeV}$. Up to $\mathcal{O}(10^4)$ and $\mathcal{O}(10^3)$ measurements are performed by employing multiple sources on the $32^3 \times 64$ and $48^3 \times 96$ ensembles, respectively. \\

\vspace{-3mm} \noindent
We obtain amplitudes for several values of current momentum, $Q^2$, in the range $1.5 \lesssim Q^2 \lesssim 7$ GeV$^2$. Multiple $\omega$ values are accessed at each simulated value of $\bv{q}$ by varying the nucleon momentum $\bv{p}$. This analysis is performed to map out the $\omega$ dependence of the Compton structure functions given in \Cref{eq:compF1,eq:compF2} for each $Q^2$ value that we study. \\

\begin{figure}[ht]
    \centering
    \includegraphics[width=.8\textwidth]{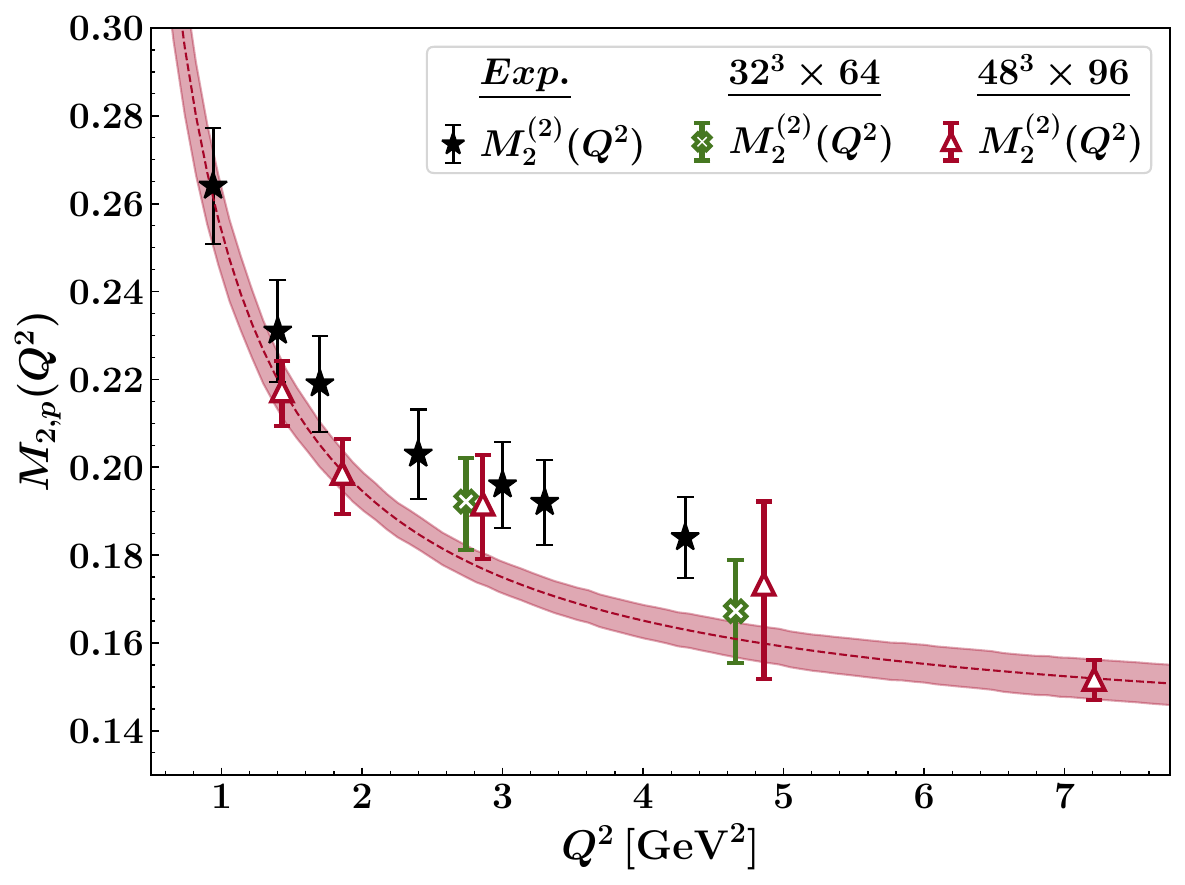}
    \caption{\label{fig:F2_moments}$Q^2$ dependence of the lowest moments of proton $F_{2}$. Filled stars are the experimental Mellin moments of $F_2$~\cite{Armstrong:2001xj}. Figure taken from~\cite{QCDSF:2022ncb}.
    }
\end{figure}

\vspace{-3mm} \noindent 
The first few Mellin moments of $F_1$, and $F_2$ are determined by performing a simultaneous fit to $\overline{\mathcal{F}}_1$ and $\mathcal{F}_2$ in a Bayesian framework at each $Q^2$ value (see Refs.~\cite{PhysRevD.102.114505,QCDSF:2022ncb} for details). We show the lowest moments of $F_{2}$ for proton in \Cref{fig:F2_moments} as a function of $Q^2$. Also shown are the experimental determinations of the Mellin moments of $F_{2}$~\cite{Armstrong:2001xj}. We see a very good agreement, although we should note that our systematics are not fully accounted for yet. \\

\vspace{-3mm} \noindent
The Compton amplitude encompasses all power corrections, therefore it is possible to estimate the leading power correction (i.e. twist-4) by studying the $Q^2$ behaviour of the moments in a twist expansion,
$M_{2,p}^{(2)}(Q^2) = M_{2,p}^{(2)} + C_{2,p}^{(2)}/Q^2 + \mathcal{O}(1/Q^4)$. Utilising only the $M_2^{(2)}(Q^2)$ moments obtained on the $48^3 \times 96$ ensemble, we study the power corrections down to $Q^2 \approx 1.5 \; {\rm GeV}^2$. Our fit for proton is shown in \Cref{fig:F2_moments}. The power corrections seem to become important below $5 \, {\rm GeV}^2$. \\

\vspace{-3mm} \noindent
Power corrections are a combination of target mass corrections, pure higher-twist terms, and the elastic contributions. These effects can be disentangled further, for instance by determining the elastic contributions from form factors~\cite{Carlson:1998gf,Melnitchouk:2001eh}, and employing Nachtmann moments~\cite{NACHTMANN1973237} to account for the target mass corrections, along with including the logarithmic evolution of the moments. \\

\vspace{-3mm} \noindent
As a first step, we incorporate the logarithmic evolution and perform a global fit to our available Compton amplitude results up to $5 \, {\rm GeV}^2$. We assume a parton distribution function-like parametric form for the $F_2$ structure function, which additionally includes the $\mathcal{O}(1/Q^2)$ power corrections,
\begin{equation} \label{eq:pdf}
    \tilde{f}_q(x,Q^2) = a_q x^{b_q} (1-x)^{c_q} \left( 1 + \frac{d_q x^{e_q} (1-x)^{f_q}}{Q^2} \right), 
\end{equation}
where $a_q$, $b_q$, $c_q$, $d_q$, $e_q$, and $f_q$ are free fit parameters, and $q=u,d$. Evaluating the integral in \Cref{eq:compomega12} by replacing $F_2(x,Q^2)$ with \Cref{eq:pdf}, we obtain,
\begin{equation}
     \frac{\mathcal{F}_2(\omega,Q^2)}{4\omega} = M_2(Q^2) \sum_{n=1}^{N} \left( A_{2n}(b,c,Q^2) + \frac{C_{2n}(b,c,d,e,f,Q^2)}{Q^2} \right) \omega^{2n-2},
\end{equation} 
where $M_2(Q^2)$ is the lowest even moment of $F_2$, and $A_{2n}$, and $C_{2n}$ are known coefficients of generalised hypergeometric series. We take $N=8$. \\

\begin{figure}[ht]
    \centering
    \includegraphics[width=.8\textwidth]{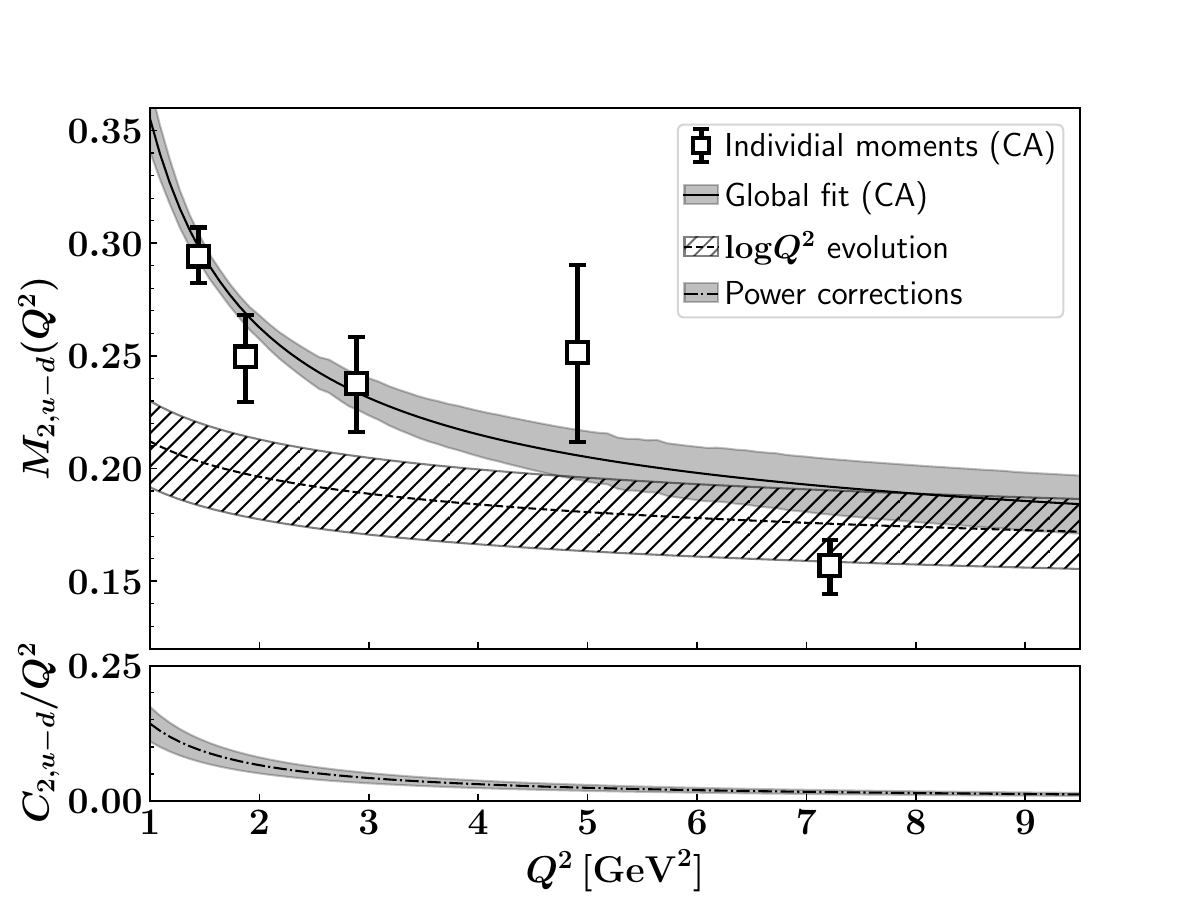}
    \caption{\label{fig:F2_umd_moments} $Q^2$ dependence of the lowest isovector moment of $F_{2}$ (solid shaded curve). We show the contributions coming from the logarithmic evolution (hatched curve) and the $1/Q^2$ term (bottom panel).
    }
\end{figure}

\vspace{-3mm} \noindent
We employ a Bayesian fitting framework, with priors of the free parameters chosen as uninformative Gaussian distributions around phenomenologically-motivated values. We fit to the $uu$ and $dd$ Compton amplitudes simultaneously, taking the correlations into account. We include the leading-order non-singlet logarithmic evolution of the moments for the fit. The resulting fit (the ``Global fit (CA)'' curve) is shown in \Cref{fig:F2_umd_moments} in comparison to the moments (squares) determined from fits to the individual Compton amplitudes at each $Q^2$. We also separate the contributions of the logarithmic evolution (hatched curve) and the power corrections (bottom panel). The power corrections start to become important below $\sim 5 \, {\rm GeV}^2$. 
\vspace{-10mm}
 \\

\begin{figure}[ht]
    \centering
    \includegraphics[width=\textwidth]{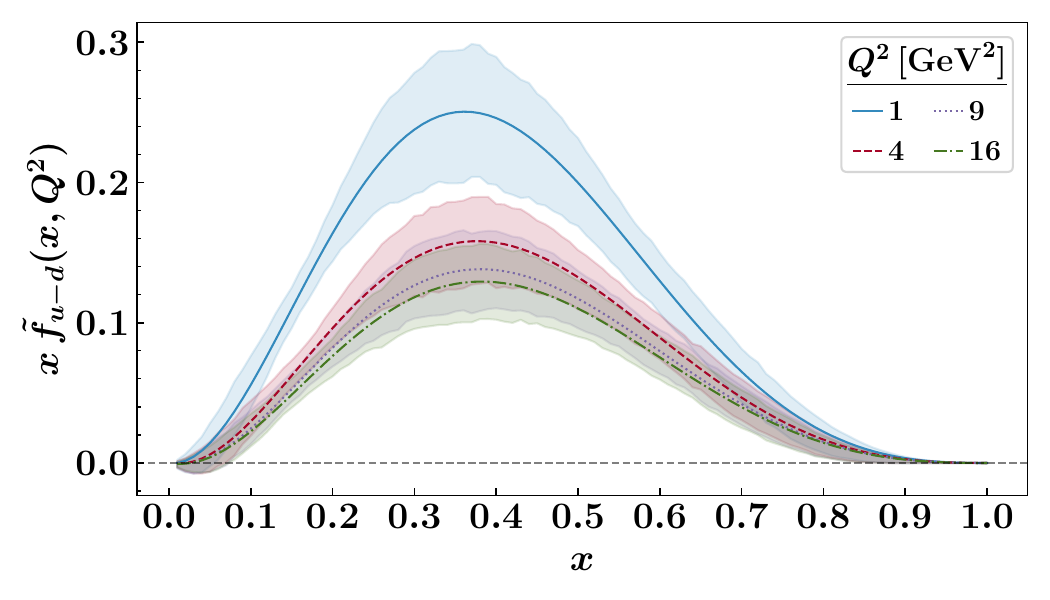}
    \caption{\label{fig:pdf_wht} $Q^2$ dependence of $x \tilde{f}_{u-d}(x,Q^2)$ including the $\mathcal{O}(1/Q^2)$ power corrections. Note that this is at the $SU(3)$ symmetric point.
    }
\end{figure}

\vspace{-3mm} \noindent
In \Cref{fig:pdf_wht}, we plot \Cref{eq:pdf} for the isovector combination at different $Q^2$ values, using the parameters determined from a global fit to the lattice Compton amplitude results at the $SU(3)$ symmetric point. Although subtle, we see that the peak shifts and the low- and high-x slopes change as we increase the $Q^2$. Once we are in the region $Q^2 \gtrsim 10 \, {\rm GeV}^2$, the variation is only due to the logarithmic evolution, hence slow. 

\section{Summary and outlook}
With the Compton amplitude approach it is possible to directly investigate the structure functions including the effects beyond leading twist. We showed the versatility of this approach by calculating the moments of transverse structure function, $F_2$, along with their $Q^2$ dependence. This allows us to study the physical power corrections in a lattice calculation. We have taken the first steps towards disentangling the logarithmic evolution, power corrections and genuine higher-twist effects by considering a global fit to our available lattice Compton amplitude results. \\ 

\vspace{-3mm} \noindent
Currently our calculations involve configurations with two different lattice spacings and volumes, all at the $SU(3)$ symmetric point. Calculations on additional ensembles that cover a range of lattice spacings and pion masses are required to fully account for systematic effects and make contact to the phenomenology.

{\small
\subsection*{Acknowledgments}
The numerical configuration generation (using the BQCD lattice QCD program~\cite{Haar:2017ubh})) and data analysis (using the Chroma software library~\cite{Edwards:2004sx}) was carried out on the DiRAC Blue Gene Q and Extreme Scaling (EPCC, Edinburgh, UK) and Data Intensive (Cambridge, UK) services, the GCS supercomputers JUQUEEN and JUWELS (NIC, Jülich, Germany) and resources provided by HLRN (The North-German Supercomputer Alliance), the NCI National Facility in Canberra, Australia (supported by the Australian Commonwealth Government) and the Phoenix HPC service (University of Adelaide). RH is supported by STFC through grant ST/P000630/1. PELR is supported in part by the STFC under contract ST/G00062X/1. KUC, RDY and JMZ are supported by the Australian Research Council grants DP190100297 and DP220103098. For the purpose of open access, the authors have applied a Creative Commons Attribution (CC BY) licence to any Author Accepted Manuscript version arising from this submission.
}
{\small
\bibliographystyle{JHEP}
\begingroup
    \setlength{\bibsep}{1pt}
    \setstretch{0}
    % \bibliography{can_dis23}

\providecommand{\href}[2]{#2}\begingroup\raggedright\endgroup

\endgroup
}
\end{document}